\documentclass[prb, showpacs, superscriptaddress, twocolumn]{revtex4}
\usepackage{graphicx}
\usepackage{amssymb}
\usepackage{subfigure}

\begin{document}
\title{The degree of 5{\it{f}} electron localization in URu$_2$Si$_2$: electron energy-loss spectroscopy and spin-orbit sum rule analysis}

\author{J. R. Jeffries}
\affiliation{Condensed Matter and Materials Division, Lawrence Livermore National Laboratory, Livermore, CA 94550, USA}
\author{K. T. Moore}
\affiliation{Condensed Matter and Materials Division, Lawrence Livermore National Laboratory, Livermore, CA 94550, USA}
\author{N. P. Butch}
\affiliation{Center for Nanophysics and Advanced Materials, Department of Physics, University of Maryland, College Park, MD 20742, USA}
\author{M. B. Maple}
\affiliation{Department of Physics, University of California, San Diego, La Jolla, CA 92093, USA}

\date\today

\begin{abstract}
We examine the degree of 5{\it{f}} electron localization in URu$_2$Si$_2$ using spin-orbit sum rule analysis of the U N$_{4,5}$ (4{\it{d}}~$\rightarrow$~5{\it{f}} ) edge. When compared to $\alpha$-U metal, US, USe, and UTe, which have increasing localization of the 5{\it{f}} states, we find that the 5{\it{f}} states of URu$_2$Si$_2$ are more localized, although not entirely. Spin-orbit analysis shows that intermediate coupling is the correct angular momentum coupling mechanism for URu$_2$Si$_2$ when the 5{\it{f}} electron count is between 2.6 and 2.8. These results have direct ramifications for theoretical assessment of the hidden order state of URu$_2$Si$_2$, where the degree of localization of the 5{\it{f}} electrons and their contribution to the Fermi surface are critical.
\end{abstract}



\maketitle

At $T_0$=17.5 K, URu$_2$Si$_2$ undergoes a second-order transition from a moderately heavy fermion paramagnetic state to a ``hidden order'' (HO) state.\cite{Palstra1985, Schlabitz1986, Maple1986}  While the transition into the HO state is characterized by anomalies in nearly every bulk, physical properties measurement, a satisfactory order parameter describing this HO state has remained elusive since its discovery over 20 years ago.  Initial neutron scattering experiments below $T_0$ detected a small antiferromagnetic moment at a commensurate wavevector, but this moment was insufficient to account for the entropy released during the HO transition;\cite{Broholm1987, Broholm1991} the small moment at ambient pressure is now generally regarded as extrinsic in nature.\cite{Niklowitz2009a}  Recent neutron scattering experiments reveal the presence of gapped spin excitations centered at an incommensurate wavevector, and the gapping of those excitations is able to account for the entropy liberated by the HO transition.\cite{Wiebe2007} In addition to the gapped spin excitation spectrum, the HO transition opens a partial gap on part of the Fermi surface.  The remainder of the partially gapped portion of the Fermi surface becomes gapped by the onset of superconductivity at $T_c$=1.5 K.\cite{Maple1986}

The HO state seems to be intimately linked with the presence of antiferromagnetic fluctuations.  With applied pressure, the HO transition temperature increases,\cite{Jeffries2007, Butch2009b} and the HO state yields to an antiferromagnetic (AFM) ground state characterized by a commensurate sublattice magnetization near 0.4 ${\mu}_B$.\cite{Amitsuka1999}  The details of this HO/AFM transition are the subject of much current debate, as the relationship between these two ground states can reveal important information about the HO state.\cite{Bourdarot2005} Applied pressure causes the partial Fermi surface gap opened by the HO transition to increase at the expense of superconductivity,\cite{Jeffries2008} which seemingly does not survive into the bulk AFM state.\cite{Hassinger2008}  High magnetic field studies have shown the HO state to be robust out to extremely large fields, and the complex field-induced phenomena associated with the HO state are only modestly affected by applied pressure.\cite{Jo2007}  Furthermore, the addition of small concentrations of Re into the URu$_2$Si$_2$ lattice suppresses the AFM correlations along with the HO state, leading to ferromagnetic fluctuations, long-range ferromagnetism, and criticality.\cite{Butch2009a}

The myriad experiments performed on URu$_2$Si$_2$ implicate a Fermi surface instability as a key characteristic driving the transition into the HO state, and the continuity between the HO state the pressure-induced AFM state suggests linked order parameters.\cite{Mineev2005}  Theoretical models invoke multipole ordering, crystalline electric fields, or magnetic contributions from {\it{f}}-electrons as potential mechanisms responsible for the occurrence of the HO state,\cite{Kiss2005, Haule2009, Elgazzar2009, Cricchio2009} and these calculations produce Fermi surfaces and nesting vectors that are converging and are in good agreement with experiment.\cite{Denlinger2001}  However, there are still pivotal parameters of theories yet to be grounded empirically. The degree of {\it{f}}-electron localization as wells as the number of {\it{f}}-electrons and their hybridization play important roles in calculations of the Fermi surface and its viable nesting vectors. To that end, we have investigated the U 5{\it{f}} states of URu$_2$Si$_2$ using EELS in a transmission electron microscope (TEM). Using the U N$_{4,5}$ edge, the spin-orbit sum rule, and comparison to other U-based materials, we show that the 5{\it{f}} states are near intermediate coupling and exhibit moderately localized behavior. These results have direct ramifications for computational assumptions used for assessment of the low-temperature HO state in URu$_2$Si$_2$.

A single crystal of URu$_2$Si$_2$ was grown via the Czochralski technique in a tetra-arc furnace with an argon atmosphere.  The crystal was annealed at 900 $^\circ$C with a Zr getter in a partial pressure of argon for 7 days.  The crystal was confirmed to be a high-quality single crystal with the Laue method as well as TEM characterization. The TEM specimens were polished down to a thickness of 150 $\mu$m, dimpled from both sides to a thickness of 30 $\mu$m using abrasive diamond suspensions, and, finally, perforated by ion milling.  A Philips CM300 field-emission-gun TEM operating at 297 keV and equipped with a Gatan image filter was used for these EELS experiments.  The specimen of URu$_2$Si$_2$ was examined both at room temperature and 8K using a liquid helium specimen holder. 

\begin{figure}[t]
\begin{center}\leavevmode
\includegraphics[scale=0.5]{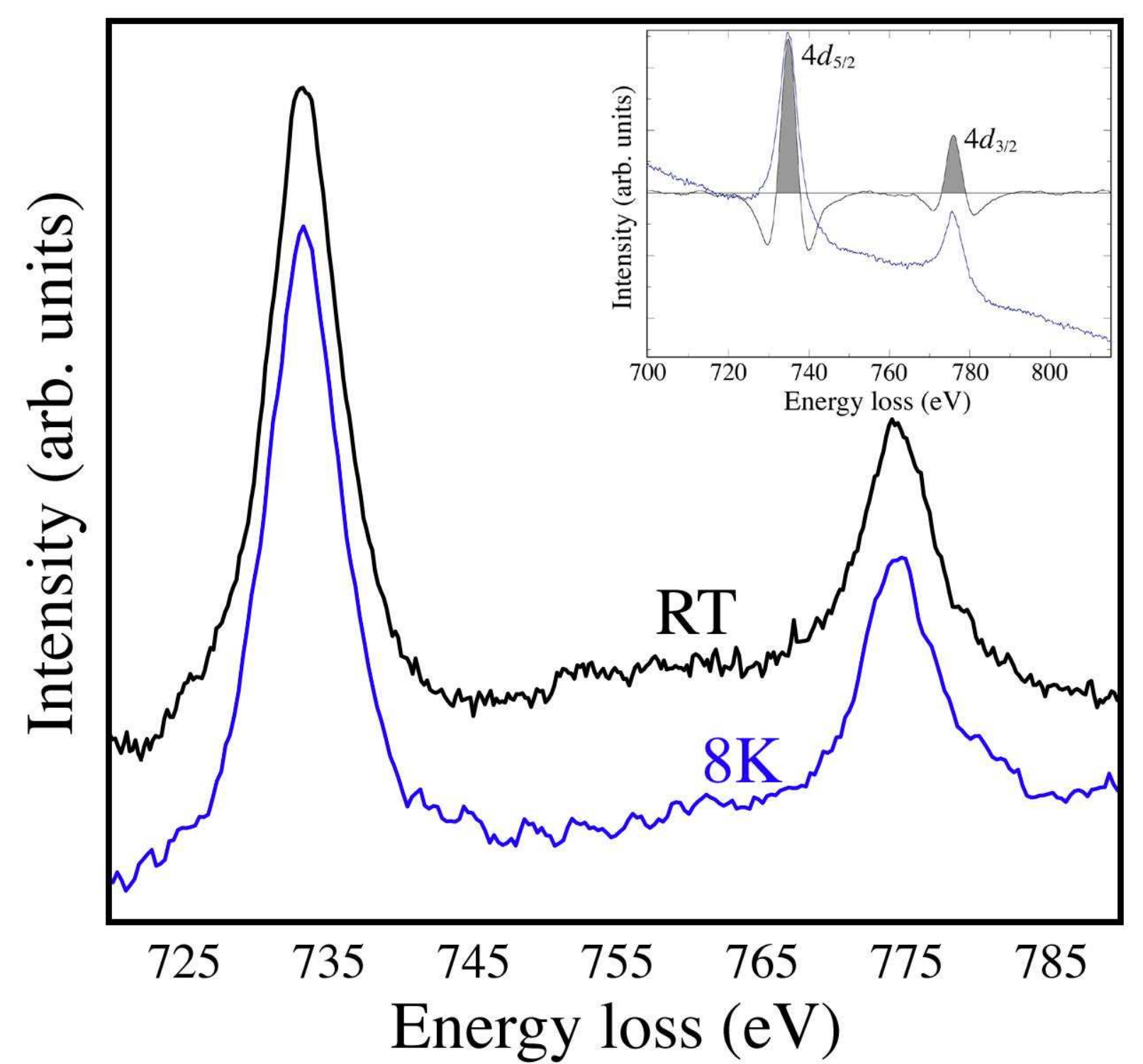}
\caption{(color online) The N$_{4,5}$ EELS spectra of URu$_2$Si$_2$ acquired in a TEM both at room temperature and 8 K. Inset: A compound graph of a raw U N$_{4,5}$ EELS spectrum and the second derivative of that spectrum illustrating how the area beneath each peak was integrated to calculate the branching ratio for spin-orbit analysis.}\label{EELS}
\end{center}
\end{figure}

The 5{\it{f}} states of actinide materials can be directly examined via electron energy-loss spectroscopy (EELS) or x-ray absorption spectroscopy (XAS).\cite{Moore2009, Moore2010}  In each process, a core {\it{d}} electron is excited above the Fermi energy to the unoccupied states, where electric-dipole selection rules allow two kinds of transitions: {\it{d}}$_{5/2}~\rightarrow~$5{\it{f}}$_{5/2,7/2}$ and {\it{d}}$_{3/2}~\rightarrow~$5{\it{f}}$_{5/2}$. The 5{\it{f}} spin-orbit interaction per hole can be directly examined for an actinide material using these excitations and the spin-orbit sum rule.\cite{Thole1988, vanderLaan1996, vanderLaan2004}  For this analysis, the branching ratio must be extracted from the spin-orbit-split core-level edges in the the M$_{4,5}$ (3{\it{d}}~$\rightarrow$~5{\it{f}}) or the N$_{4,5}$ (4{\it{d}}~$\rightarrow$~5{\it{f}}) spectra. Analyzed spectra were composed of a sum of 12 to 15 individual spectra, enhancing signal to noise. Background removal for the spectra in Fig. \ref{EELS} was performed with an inverse power-law extrapolation.  For branching ratio analysis, the second derivative of the raw spectra was calculated and the area beneath the N$_5$ (4{\it{d}}$_{5/2}$) and N$_4$ (4{\it{d}}$_{3/2}$) peaks was integrated above zero.  This technique, which is illustrated in the inset in Fig. \ref{EELS}, provides optimal signal to noise in the spectrum and circumvents the need to remove the background intensity with an inverse power-law extrapolation as described above. The branching ratio, $B~=~$I(N$_{5}$)/[I(N$_{5}$) + I(N$_{4}$)], was obtained by integrating the area under the N$_5$ and N$_4$ peaks, yielding I(N$_5$) and I(N$_4$), respectively.

The N$_{4,5}$ EELS spectra for URu$_2$Si$_2$ are shown in Fig. \ref{EELS} for both room temperature and 8K.  Each spectrum displays two ``white-line'' peaks, a strong N$_5$ (4{\it{d}}$_{5/2}~\rightarrow~$5{\it{f}}$_{5/2,7/2}$) and a smaller N$_4$ (4{\it{d}}$_{3/2}~\rightarrow~$5{\it{f}}$_{5/2}$).  The branching ratio for the EELS spectra are 0.719 and 0.723 for room temperature and 8K, respectively (see Table \ref{Table} for comparison with other U-based compounds). 

\begin{table}[b]
\caption{The measured branching ratios, $B$, of the N$_{4,5}$ EELS/XAS spectra and the expectation values of the 5{\it{f}} spin-orbit interaction per hole, $<w^{110}>$/(14-n$_f$)-$\Delta$, for $\alpha$-U, URu$_2$Si$_2$ at room temperature (RT) and 8K, US, USe,and UTe. Uncertainties in the last digit(s) of the quantities determined from EELS are given in parentheses with the values.}\label{Table}
\begin{center}
\begin{tabular}{l c c}
\hline
\hline
Material &~~~~~$B$~~~~~& $~~<w^{110}>$/(14-n$_f$)-$\Delta$~~ \\
\hline
$\alpha$-U\footnotemark[1] 	& 0.686(2) 	& -0.215(5) \\
US\footnotemark[2] 			& 0.70		& -0.250 \\
URu$_2$Si$_2$ (RT)~~		& 0.719(5)	& -0.298(13) \\
URu$_2$Si$_2$ (8 K)~~		& 0.723(5)	& -0.308(13) \\
USe\footnotemark[2]			& 0.73		& -0.325 \\
UTe\footnotemark[2]			& 0.74		& -0.350 \\
\hline
\hline
\end{tabular}
\end{center}
\footnotetext[1]{From K. T. Moore, {\it{et al.}}\cite{Moore2006}}
\footnotetext[2]{From T. Okane, {\it{et al.}}\cite{Okane2004, Okane2008}}
\end{table}

Using the atomic spectral calculations and sum-rule analysis of van der Laan and Thole\cite{vanderLaan1996} for the URu$_2$Si$_2$ EELS spectra, we can examine the transitions in detail.   For the {\it{f}} shell, the expectation value of the angular part of the spin-orbit parameter is 

\begin{eqnarray*}
<w^{110}>&=&\frac{2}{3}<l{\cdot}s>~=~n_{7/2}-\frac{4}{3}n_{5/2},
\end{eqnarray*} 

\noindent where $n_{7/2}$ and $n_{5/2}$ are the electron occupation numbers for the angular-momentum levels j = 7/2 and 5/2.  Thus, $<w^{110}>$ reveals the proper angular momentum coupling scheme for a given material.  For the {\it{d}}~$\rightarrow$~{\it{f}} transition, the sum rule gives the spin-orbit interaction per hole as

\begin{eqnarray*}
\frac{<w^{110}>}{14-n_f}-{\Delta}&=&-\frac{5}{2}\left(B-\frac{3}{5}\right),
\end{eqnarray*}

\noindent where $B$ is the measured branching ratio for the experimental EELS spectra, $n_f$ is the number of electrons in the {\it{f}} shell, and $\Delta$ represents the small correction term for the sum rule that is calculated using CowanÕs relativistic Hartree-Fock code.\cite{vanderLaan2004, Cowan1981}

\begin{figure}[t]
\begin{center}\leavevmode
\includegraphics[scale=0.45]{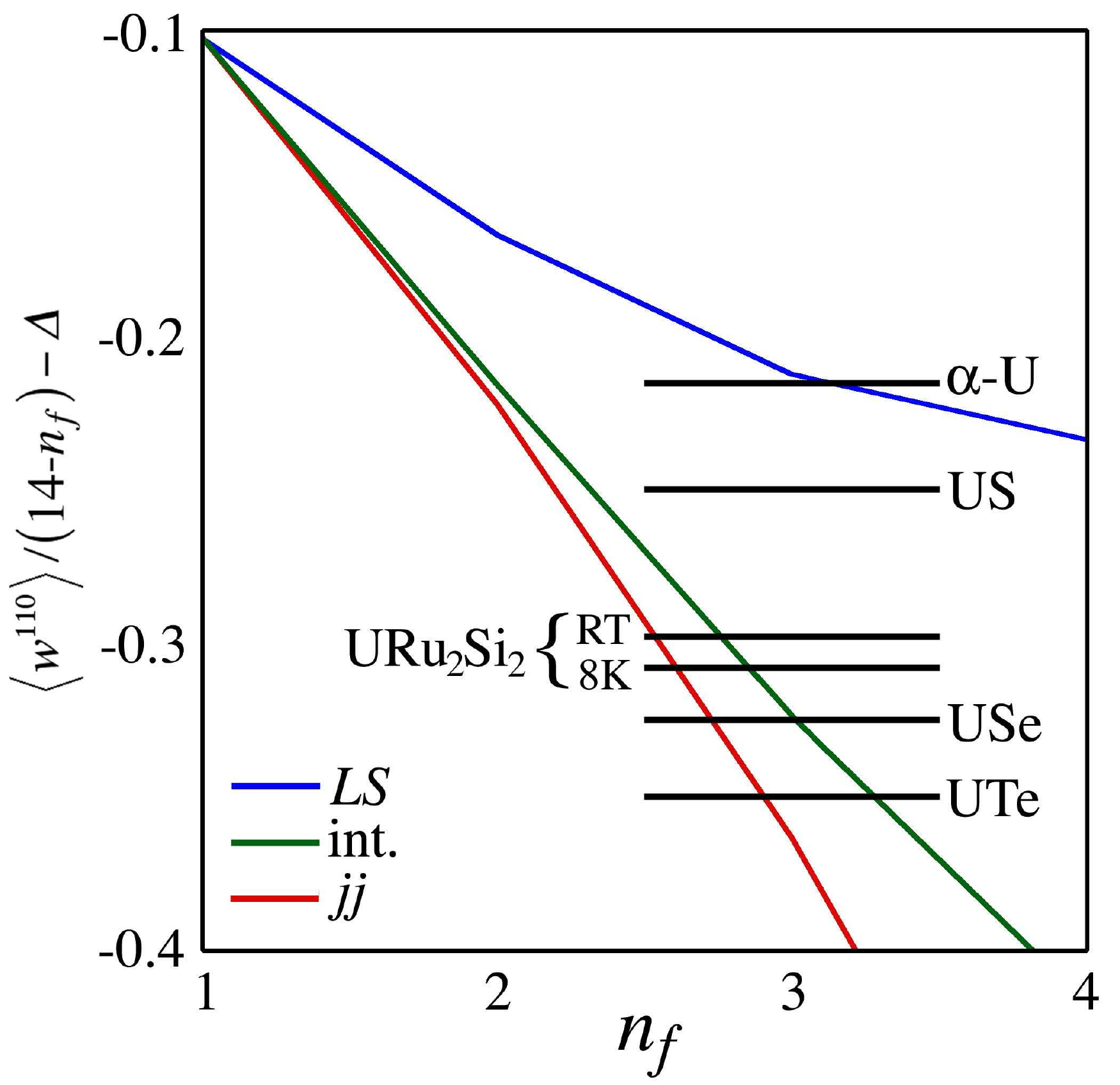}
\caption{(color online) Ground-state spin-orbit interaction per hole, $<w^{110}>$/(14-n$_f$)-$\Delta$, as a function of the number of 5{\it{f}} electrons (n$_f$). The three theoretical angular momentum coupling schemes are shown: {\it{LS}} (blue curve), intermediate (green curve), and {\it{jj}} (red curve). Data from the experimentally measured branching ratios of URu$_2$Si$_2$ are indicated by horizontal black lines.  Also included are EELS results for $\alpha$-U\cite{Moore2006} and XAS results for US, USe, and UTe.\cite{Okane2004, Okane2008}}\label{BranchingRatio}
\end{center}
\end{figure}

The results of the spin-orbit analysis of the N$_{4,5}$ EELS spectra are plotted as horizontal lines in Fig. \ref{BranchingRatio}, since the number of electrons in the 5{\it{f}} shell is not an output of the analysis. In addition to the present URu$_2$Si$_2$ data, spin-orbit analysis was performed on EELS results for $\alpha$-U metal\cite{Moore2006} as well as XAS results for US, USe, and UTe.\cite{Okane2004, Okane2008} Also included in the graph are the {\it{LS}}, intermediate, and {\it{jj}} coupling mechanisms for the angular momenta, from atomic calculations,\cite{vanderLaan1996} plotted against $n_f$ as a blue, green, and red line, respectively. Immediately noticeable is that the results of the spin-orbit analysis form a continuum from $\alpha$-U near the {\it{LS}} curve to UTe near the {\it{jj}} and intermediate curves. It is the location of the branching ratio of URu$_2$Si$_2$ within this continuum of angular momentum coupling mechanisms that yields insight into the nature of the 5{\it{f}} states of URu$_2$Si$_2$.

The branching ratio and spin-orbit sum rule are sensitive to the degree of electron localization, meaning they can be used to examine the relative degree of 5{\it{f}} itinerancy (or, conversely, localization).\cite{vanderLaan2004, Moore2007} This arises due to the fact that once a state becomes localized it exhibits intermediate coupling due to competition between spin-orbit and exchange interaction.  An example of this is found in the 5{\it{d}} transition metals, which have delocalized and bonding 5{\it{d}} states. For those metals an {\it{LS}}-coupling mechanism is appropriate; however, once these metals form dioxides, the 5{\it{d}} states localize and adopt intermediate coupling. Furthermore, in the case of the rare earth metals with localized 4{\it{f}} states, exchange interaction dominates the spin-orbit interaction, positioning the intermediate coupling mechanism very near the {\it{LS}} limit. The closeness of intermediate coupling of the rare earth 4{\it{f}} states to the {\it{LS}} limit allows {\it{LS}} coupling to work well for calculations even though the 4{\it{f}} states are governed by intermediate coupling. The light actinide metals have a strong spin-orbit interaction that dominates the exchange interaction once the 5{\it{f}} states are localized via chemical doping, oxidation, etc. This, in turn, positions intermediate coupling close to the {\it{jj}} limit, as shown in Fig. \ref{BranchingRatio}. The change in angular momentum coupling mechanism to intermediate coupling upon localization is precisely what allows the branching ratio and spin-orbit sum rule to assess the degree of 5{\it{f}} itinerancy or localization.

Examining Fig. \ref{BranchingRatio}, an {\it{LS}} mechanism is observed for $\alpha$-U metal with $n_f\approx$3. The 5{\it{f}} states of $\alpha$-U are clearly itinerant, as evidenced by the equation of state of U, which exhibits no structural change upon compression to 100 GPa in a diamond anvil cell.\cite{Akella1997}  Likewise, the 5{\it{f}} states of US are primarily itinerant in nature. This is shown by the spin-orbit analysis in Fig. \ref{BranchingRatio}, which falls between the {\it{LS}} and intermediate coupling curves when band structure calculations that show $2.6{\lesssim}n_f{\lesssim}$2.9 for US are taken into account.\cite{Trygg1995, Yamagami1998} The delocalized character of the 5{\it{f}} states in US are further supported by recent soft x-ray photoemission spectroscopy.\cite{Takeda2009} The 5{\it{f}} states of USe are localized to a greater extent and reside near the intermediate coupling curve for $2.8{\lesssim}n_f{\lesssim}$3.1. The 5{\it{f}} states of UTe are almost entirely localized,\cite{Lander1990} and, accordingly, the spin-orbit analysis of UTe is lower than USe, forming an end limit of the relative continuum of 5{\it{f}} electron localization.\cite{Note}

Now considering the spin-orbit analysis of URu$_2$Si$_2$ in relation to the other U materials, we see that the 5{\it{f}} states are moderately localized at both room temperature and 8K, falling on the intermediate coupling curve for $n_f$ between approximately 2.7 and 2.8. The measured branching ratio exhibits very little temperature dependence, suggesting that the 5{\it{f}} electrons remain equivalently localized regardless of the Fermi surface rearrangement associated with the onset of the HO state.  There is no 5{\it{f}} electron occupation that brings URu$_2$Si$_2$ near the {\it{LS}} coupling limit. These results show that the 5{\it{f}} states of URu$_2$Si$_2$ are localized enough to exhibit intermediate coupling, but are not as localized as those of USe or UTe.  As the coupling scheme is unphysical for values of $n_f$ to the left of the {\it{jj}} limit, the measured branching ratio implies an occupancy of $n_f~{\gtrsim}~$2.5.  This suggests that treating the U ions in URu$_2$Si$_2$ as tetravalent ions is most likely incorrect.  In all likelihood, the U ions exhibit a superposition of oxidation states including at least tetravalent and trivalent ions, with the possibility of a divalent component as well.

These results can be used to constrain theoretical models of URu$_2$Si$_2$.  The measured branching ratio provides a direct probe of the spin-orbit-coupling strength and an estimate of the occupancy of {\it{f}}-electron valence states, both of which should be tunable parameters within models derived using density-functional theory (DFT) or dynamical mean-field theory (DMFT).  While LDA calculations alone have been shown to overestimate the branching ratios actinide-bearing compounds, an LDA+DMFT approach, where the correlated {\it{f}}-states are treated within the DMFT impurity solver, has been shown to yield branching ratios in good agreement with experiment.\cite{Shim2009}  As such, extension of these techniques to URu$_2$Si$_2$ should hold promise of yielding results concordant with the measured branching ratio, while providing a foundation for exploring the other aspects of the electronic structure that give rise to the HO state and its associated properties in URu$_2$Si$_2$.
\newline

Sample synthesis was supported by the U.S. Department of Energy under Research Grant DE-FG-02-04ER46105.  JRJ and KTM are supported by the Science Campaign at Lawrence Livermore National Laboratory.  NPB is supported by the Center of Nanophysics and Advanced Materials.  MBM is supported by the National Nuclear Security Administration under the Stewardship Science Academic Alliance program through the U.S. Department of Energy under Grant No. DE-FG52-06NA26205.  Lawrence Livermore National Laboratory is operated by Lawrence Livermore National Security, LLC, for the U.S. Department of Energy, National Nuclear Security Administration under Contract DE-AC52-07NA27344.

\end{document}